{ **Author** *Rainer Rehak*

**RESEARCH ARTICLE**

# ARTIFICIAL INTELLIGENCE FOR REAL SUSTAINABILITY

## What is Artificial Intelligence and Can it Help with the Sustainability Transformation?

The discussion about the disruptive possibilities of a technology called artificial intelligence (AI) is on everyone's lips. Companies and countries alike are running multi-billion-dollar research programmes to ensure they do not miss out on the global innovation hunt. Among many other applications, AI is also supposed to aid the large-scale changes needed to achieve sustainable societies. To assess those possibilities, this article briefly explains, classifies, and theorises AI technology and then politically contextualises that analysis in light of the sustainability discourse.

Like few other technologies, AI is surrounded by almost magical promises **(Weizenbaum, 1993)**, aligning well with the long-standing narrative of the imminent digital revolution described by a seemingly recent quote from Hubert L. Dreyfus **(1972:xxvii)**: «*Every day we read that digital computers play chess, translate languages, recognize patterns, and will soon be able to take over our jobs. In fact, this now seems like child's play*». The fact that the quote is from 1972 shows how necessary a nuanced analysis of the ‹AI field› is.

Originally, the term ‹AI› referred to a field of computer science, which initially had the dry name ‹automata theory›. However, computer scientists John McCarthy and Marvin Minsky believed the term ‹artificial intelligence› would be more fitting, particularly in terms of the popularity and fundability of the field, so they renamed it for a workshop in 1955. AI is not a single technology in itself but rather a diverse field within computer science. For a long time, AI systems explicitly represented knowledge or ‹symbolised› it, enabling logical reasoning and decision trees **(Bonsiepen, 1994)**. These symbolic approaches allow formal conclusions such as ‹all sweet things are sticky and jam is sweet, therefore jam is sticky›.

Later in the 1980s, the so-called sub-symbolic approaches emerged, such as artificial neural networks (ANNs), genetic algorithms, or other statistical and heuristic approximations. These approaches require pre-configuration with large amounts of data in many iterations, often called ‹training›, in which knowledge is only implicitly represented and difficult to verify **(Mitchell, 1997; Rehak 2021a)**. Current image recognition, large language models, and translation systems work in this way.

### THREE TYPES OF AI

In the academic AI discourse, two types of AI are usually distinguished: strong AI and weak AI. However, I would like to add a third type, which I call Zeitgeist AI.

Strong AI, also known as Artificial General Intelligence (AGI), refers to a system that possesses general and flexible intelligence, can ask questions, and exhibits genuine creativity, perhaps even consciousness. Such a system could act independently, potentially have its own goals, and would, therefore, need to assume responsibility for its agency. This type of AI exists only in the realm of science fiction, and there are no signs in technical AI research that this will change in the foreseeable future **(Kurzweil, 2005; Rehak, 2021a)**.

```
///<quote>
   It cannot change domains
or set its own goals;
it is a tool,
albeit a very complex one.
///</quote>
```

Weak AI, also known as Artificial Narrow Intelligence (ANI), refers to a system that can only perform narrowly defined, highly specialised, and domain-specific tasks. It cannot change domains or set its own goals; it is a tool, albeit a very complex one. This category includes systems that recognise patterns (visual and acoustic object recognition or data analysis such as optimising resource use, e.g., electricity and water consumption) or automatically perform other domain-specific tasks with clear objectives (e.g., playing Go, producing derived text or images). All current AI systems fall into this category, including the current large language models **(Bender et al., 2021)**.

Thirdly, I would like to define Zeitgeist AI as a discursive phenomenon where political, societal, and even academic actors refer to ‹AI› when they actually mean anything related to complex digital technologies such as algorithms, big data, software, programmes, computer systems, automation, IT, actual AI, statistics, and even digitalisation in general **(Council for Social Principles of Human-centric AI, 2019)**. With such a vague AI concept, serious and fruitful AI debates are difficult, which is why they regularly need to be reined in **(Butollo, 2018; Eyert et al., 2020)**.

### PRECISE LANGUAGE

Moreover, great caution is needed in the choice of language when discussing AI, as many of the prevailing technical terms historically referenced human activities and abilities but should not be understood as analogies. The terms ‹act›, ‹decide›,

```
///<quote>
   Incorrect terms evoke
false associations.
///</quote>
```

‹recognise›, ‹understand›, ‹(self-)learn›, ‹know›, ‹train›, ‹autonomy›, ‹predict›, and even ‹intelligence› are highly misleading. Incorrect terms evoke false associations, fuel unfounded technology fictions, and imply nonsensical or even (societally) harmful applications **(Weizenbaum, 1978)**. Appropriate terms have been suggested, e.g.,

‹move›, ‹execute›, ‹detect›, ‹conform to expectations›, ‹dynamic configuration›, ‹data/information›, ‹pre-configuration›, ‹automation›, ‹projection›, and ‹complex data/information processing› **(Rehak, 2021a; Olson, 2021)**. Such terms are especially relevant in interdisciplinary contexts or in science communication.

### WHAT AI CANNOT DO (WELL)

AI systems can effectively do specific tasks that have clear rules, adequate models, specific goals, and suitable data available. These tasks include predictive technical maintenance (e.g., for rotating parts), resource consumption optimisation (e.g., water usage in agriculture, energy consumption in data centres), voice/image detection (e.g., speech, landmarks, and animals), and speech/image synthesis. Moreover, AI can be used to search any data for patterns (compartmentalisation, clustering, etc.). There are also impressive generative AI applications in the fields of image, language, and music, but they do not solve specific tasks and are so far of explorative value.

```
///<quote>
    Many of the characteristics
attributed to AI
are often classic computer
science methods
or even just human labour
in the Global South.
///</quote>
```

Many of the characteristics attributed to AI are often classic computer science methods **(Narayanan, 2019)**, precisely Zeitgeist AI, or even just human labour in the Global South **(Solon, 2018)**. But to discuss the potential of AI properly, we need to differentiate what we are talking about. Two insightful examples:

1 The core of automated driving (AD) is not AI, as AI is so far only responsible for image recognition (e.g., traffic sign detection), the rest is not AI; therefore funding AI would not necessarily improve AD.

2 AI systems cannot generate predictions per se. What AI can do is statistically analyse past data and calculate a mathematical projection from it. But whether the calculated result is a meaningful prediction depends heavily on the subject area **(Dreyfus, 1972)**. Weather data is fundamentally different from social data **(Lopez, 2021)**. Thus, ‹predictions› in the social domain only work when social physics is assumed, which is highly controversial in theory and practice **(Eyert and Lopez, 2023)**. Not only in the crime domain, e.g., predictive policing or recidivism, have such predictive attempts generally failed. So, we still have to understand the subject area before applying AI.

Even if the area is mathematically well-understood, we, as a society, often do not even want to make correct ‹unbiased› predictions based on the past as the basis for our actions. A purely mathematically justified credit allocation based on income, for example, would, if correctly applied, simply reproduce the gender pay gap and systematically grant lower loans to women. In this case, mathematically correct results would be unfair, and fair results would be mathematically incorrect **(Eyert and Lopez, 2021; Mühlhoff, 2020)**. To put it vividly: Making predictions with AI is like driving a car while looking exclusively in the rear-view mirror. Despite great anti-bias work in the field, there is a principal limit regarding the neutrality and fairness promise of AI **(Kleinberg et al., 2017)**.

## AI AND DIGITALISATION AS ORGANISATIONAL TOOLS

In light of the reflections on the possibilities of AI, it becomes clear that the commonly raised topic of ‹human versus machine› in the context of AI is a pseudo-problem only present in science fiction. Currently existing and conceivable AI has no personal goals or motivations – even when appearing in robotic form – and must therefore be understood as a complex tool.

```
///<quote>
    Tools are used by actors
to pursue interests
and objectives, potentially
against other organisations
or individuals.
///</quote>
```

However, viewing AI as a powerful tool necessitates expanding the scope of analysis from focusing solely on specific AI techniques, as interesting and unique as they may be, to the organisations that develop, implement, and disseminate them **(Mühlhoff, 2020)**.

Tools are used by actors to pursue interests and objectives, potentially against other organisations or individuals.

Zeitgeist AI (including AI) is always an extension of an organisation. If anywhere, AI conflicts arise along the line of ‹organisation versus organisation›, which is why the interests of the involved actors should always be at the centre of AI analyses **(Marx, 2023)**. So, if an organisation's interest does not include sustainability, AI will not be used for that (or run into the rebound effect).

This organisational view is especially important and specific to AI since AI generally encompasses data-intensive technologies and, thus – the comparison with nuclear power comes to mind – has a power-centralising effect. The fact that large companies make their AI frameworks and services freely available does not change the fact that AI can have little real benefit without the appropriate and immense data foundations. AI is, therefore, just the latest development of digital feudalism with only a few large AI providers renting out their service.

## AI FOR SUSTAINABILITY

It is, in principle, desirable when AI is applied for sustainability, ‹societal and ecological well-being› (EU High-level expert group on artificial intelligence, nd.) or for nature conservation **(Grundgesetz, Art. 20a)**. However, a holistic consideration must be made for every thoughtful use of technology, especially for such resource-intensive and centralising technologies as those covered by the umbrella term AI. The whole life cycle of AI must not create a large(r) ecological footprint elsewhere on the planet **(van Wynsberghe, 2021)**. Otherwise, the AI application itself would, despite good intentions, contribute to destroying our livelihoods. Consequently, a net positive benefit must always be sought, even if this can sometimes hardly be evaluated conclusively.

```
///<quote>
    Will new data from AI
close an information gap
and therefore allow action?
///</quote>
```

There are many examples of how specific AI systems can be used concretely for conserving resources, biodiversity, and nature **(BfN, 2023)**, and with a broad understanding of sustainability, additional applications can be added **(Rehak, 2021b)**. In those areas, good results lie ahead. However, if the AI promise of a sustainability game changer is to be fulfilled, a sincere litmus-test questions must always be asked: Is AI the best solution for moving forward on a given problem? Do we currently know too little about the exact number of certain insects, the best way to park cars in cities, or the power intake of data centres? Will new data from AI close an information gap and therefore allow action?

If yes, then we should go ahead. But if the answer is that we already know enough regarding the given domain, then applying AI just uses up vital resources, diverts political focus, and eventually acts as an excuse for inaction while time is running out. Finally, an AGI with true intelligence would probably recommend we quickly do many things we already know **(Lem, 1981)**, from 100% renewable energy use and bike cities to consequent decolonisation; so why not take a shortcut and start doing them already?

}

**ABOUT THE AUTHOR**

/// **Rainer Rehak** is a researcher at the Weizenbaum Institute for the Networked Society in the area of systemic IT security, technological sustainability, and societal data protection. He serves as an expert for parliaments and courts, is active in the Forum of Computer Scientists for Peace and Societal Responsibility (FIfF), and co-initiated the ‹Bits & Bäume› conference for digitalisation and sustainability.